# FOULING IN THIN FILM NANOCOMPOSITE MEMBRANES FOR POWER GENERATION THROUGH PRESSURE RETARDED OSMOSIS

*Arvin Shadravan*[*], *Texas A&M University, 77843, College Station, TX, United States*

*Pei Sean Goh, Universiti Teknologi Malaysia (UTM), 81310, Johor, Malaysia*

*Ahmad Fauzi Ismail, Universiti Teknologi Malaysia (UTM), 81310, Johor, Malaysia*

*Mahmood Amani, Texas A&M University at Qatar, 23874, Doha, Qatar*

*\*Corresponding Author: Email address: Arvinshadravan@tamu.edu*

## Abstract

Osmotic energy (or salinity-gradient energy) is the energy released when water with different salinities is mixed, such as rivers and oceans. By employing a semipermeable membrane to control the mixing process, the osmotic pressure gradient energy can be generated in terms of electrical power via pressure retarded osmosis (PRO) without causing adverse environmental impacts. This work presents the fabrication of thin film nanocomposite (TFN) membranes which are customized to offer high flux in forward osmosis (FO) and high osmotic power in PRO. In this study, the TFN membrane was fabricated by forming a polyamide thin film on the polysulfone substrate through the interfacial polymerization process. One of the challenges in this process is the fouling of PRO membranes. Fouling is one of the major characteristics that results in the decline in the water flux of the membrane. The hydraulic pressure during PRO processes is less than RO processes so membranes that are used for PRO are less likely to foul. Experiments show that TFN membranes are more tolerant of fouling than TFC membranes because of the nanomaterials which has higher surface hydrophilicity. The structure of the membrane is a very significant characteristic that has an influence on fouling. Especially, the structure of porosity that is coherent to the thickness. Membrane adjustment plays a key role in the trend of fouling. Zeolite loading in the polyamide layer was in the range of 0.05-0.3 wt%. Field emission scanning electron microscope (FESEM) and atomic force spectroscopy (AFM) studies indicated that zeolite was successfully embedded in the polyamide layer of the membrane. The lowest contact angle was obtained at 61.08° when 0.1wt% of zeolite was used. Similarly, after the certain cleaning strategies the 0.1wt% zeolite modified membrane resulted in higher water flux in both RO (9.66 L/m$^2$h) and FO (5.8 L/m$^2$h) experiments. TFN membrane incorporated with zeolite exhibits power density of 1.73 W/m$^2$ and retains its desired resilience to foulants for osmotic power generation.

## Introduction

Pressure retarded osmosis (PRO) is an emerging membrane process uses for harvesting energy and desalination applications. According to osmotic pressure, river water crosses in the same way as osmotic pressure due to dilution of draw solution. In addition, the low concentration solution will be depressurized in a turbine for energy production. Some of the advantages of PRO are as follows: First, the operation of PRO can function 24 hours a day. Secondly, the speed of the wind and the radiation of the sun has no influence on PRO. The third benefit is having a minor footprint and the last is the facilitated scale-up. In addition, if PRO and RO brine matches with each other then it will be good for RO brine because RO brine can be used as both solutions with high concentration and low concentration such as river water and seawater [1]. This characteristic is one of a kind for PRO processes. Moreover, wastewater can be used through PRO as a feed solution. Otherwise, PRO has some limitations that have not been ready yet commercially [2]. PRO membranes have a significant role in delivering power density. Due to high salinity in draw region leads to lower water flux which in turn reduce the production of power density. It is mainly due to the adsorption of solutes on the membrane surface. Thus, the PRO membrane should be of thin, hydrophilic and resistant to salt adsorption [3,4]. Moreover, the salt concentration of the seawater is very effective for generating energy while through PRO function the power that were generated after pumping and pre-treatment was 0.2 kWh/m$^3$.

One of the challenges is the fouling of PRO membranes. Fouling is one of the major characteristics that results in the decline in the water flux of the membrane [5]. The hydraulic pressure during PRO processes is less than RO processes so membranes that are used for PRO are less likely to foul. The structure of the membrane is a very significant characteristic that has an influence on fouling[6]. Especially, the structure of porosity that is coherent to the thickness. Membrane adjustment plays a key role in the trend of fouling [7].

The improvement of the osmotic characterization of the semi-permeable membrane is still one of the significant obstacles through PRO. One of the major issues is related to the low water flux that has consequently resulted in unpromising osmotic power [8]. The desired characteristic of PRO-based on the TFC membrane should be of a highly hydrophilic and thin layer. The conventional membranes have high water flux with lower salt rejection properties. It increases the fouling in PRO operation which ultimately deteriorates the harvesting of energy in longer periods. Thin-film nanocomposite (TFN) membranes are widely used in recent years for desalination applications. It is mainly due to the achievement of high flux without compromising the salt rejection [9]. In this regard, nanomaterials such as metal oxides and carbon have gained more attention to improve membrane performance. It can improve the characteristic properties such as hydrophilicity and strength of the material. Hydrophilicity is the main required property in attaining the higher water flux[10]. The conventional TFC RO membranes are dense skin layer, which aids in high salt retention and lower flux. Conversely, the PRO membrane should be of thin skin for the achievement of high flux. PRO-based on the mechanism of the osmotic pressure gradient. Thus, the design of the TFN membrane is the prerequisite for achieving high-power density [11]. Moreover, introducing fillers that are inorganic to the polyamide layer has shown impressive characteristics in developing TFN membrane resilience connected with resistance of the chlorine and fouling. Although there are impressive superior accomplishments, some challenges came across during the fabrication of the TFN membranes [12]. More research in this area is still needed to develop the TFN membrane with greater performance efficiency, reliability, and stability for industrial implementation. The method to improve TFN membrane fabrication can be achieved modification of hydrophilic nanofillers [13]. These can help to develop the defect-free which is organic and the fabrication of nanofillers layer containing PA/inorganic. Subsequently, the usage and the selection of nanofillers in the fabrication of TFN membranes depend on the characteristics of the feed [14].

## Experimental

During the period of osmosis process, permeate water rapidly diluted draw solution, when the feed solution is being concentrated. The pressure which is hydrostatic is entitled to osmotic pressure such as water transportation passing through the membrane while it is functionalized to draw solution. At any step while ΔP can be between zero to Δπ (Osmotic pressure), water can flow into the saline water because Δπ is still larger than the pressure which is ΔP. This process is entitled to PRO [15]. In order to calculate the water flux through PRO process we need to calculate $J_w$ which is the sign of water flux and is shown by the following equation [16]:

$$J_w = A_w(\Delta\pi - \Delta P) \tag{1}$$

In this equation $A_w$ is membrane permeability coefficient (L/m²h.bar) and ΔP is differential feed pressure (bar) and Δπ is the osmotic pressure (bar).

## Materials and Methods

The chemicals and materials utilized in this research are as follows: A commercial UF PS membrane support (PS35, Nanostone Flat Sheet Membrane) was purchased from Sterlitech Corporation. N-Hexane was obtained from Fisher Scientific. TMC, MPD, Zeolite (Molecular Sieves, 3Å), Sodium Chloride were purchased from Sigma Aldrich. Prior to TFC membrane preparation, commercial polysulfone membrane was soaked in RO water for an overnight. Interfacial polymerization technique was adopted for the fabrication of polyamide layer membrane fabrication. The schematic of interfacial polymerization is shown in Figure 1. Firstly, dried PSF substrate was mounted on a frame, 2 wt% of aqueous monomer solution of MPD was poured onto a membrane for 1 minute to form the amine layer over the substrate. Then, the excess MPD solution was drained and removed using the rubber roller. Subsequently, 0.1 wt% of TMC solution in n-hexane solution was poured onto the amine-catalyzed PSF membrane. It was allowed to react for 50 sec to form a thin polyamide layer over the commercial PSF membrane. Immediately, the thin film was allowed to get bound in substrate by keeping in an oven at a temperature of 60˚C for 5 min [17,18,19]. Finally, the reaction was terminated by storing the membranes in distilled water. For TFN membranes, the molecular sieves were dispersed using ultrasonicator in organic TMC/n-hexane solution. The above mentioned were also followed for the preparation of TFN membranes. The molecular sieves were loaded in three different fractions from 0.05 to 0.3 and labeled as Zt TFN 0.05, Zt TFN 0.1, Zt TFN 0.3 [20,21].

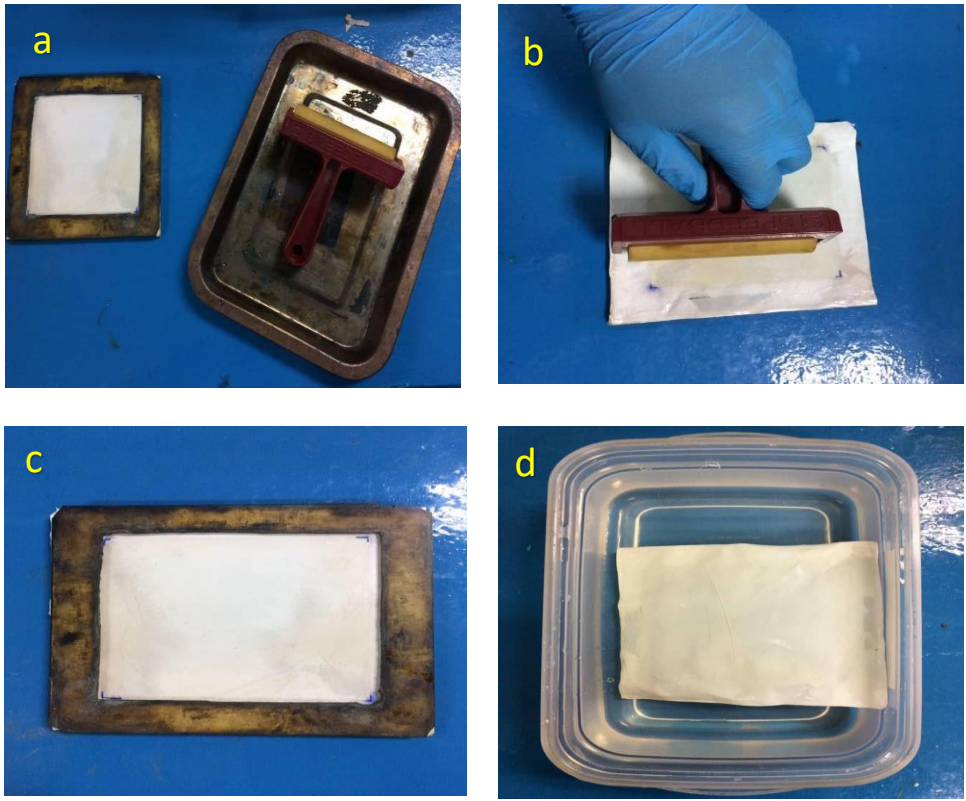

**Figure 1.** Schematic of interfacial polymerization of PA-TFC membrane. (a) PS35 commercial membrane and TFC utensils, (b) MPD treatment by roller, (c) treatment of membrane by TMC and (d) immersion of TFC membrane in RO water

## FO and PRO Filtration Studies of Membranes

Figure 2 shows the lab-scale FO and PRO cross-flow cell test setup. The feed and draw solution of RO and PRO experiments are as RO water and 2M NaCl solution respectively. The feed and draw solution were pumped through the membrane using peristaltic pump at the speed of 300 rpm. Digital weighing balance was used to monitor the draw solution tank. Feed was continuously monitored using conductivity meter [22].

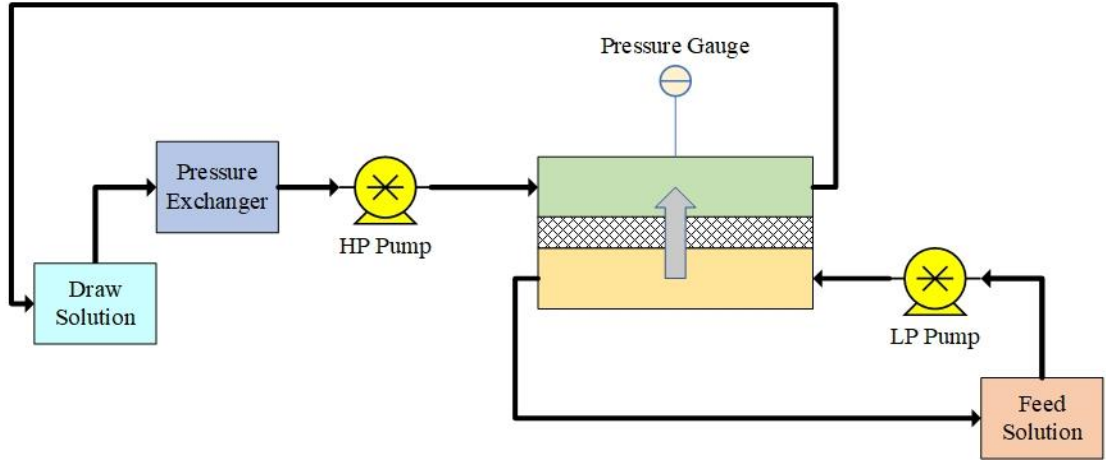

**Figure 2.** FO-PRO cross flow cell experimental setup

In FO modes, the active PA is facing towards the feed solution and draw solution is circulated over the bottom layer. Feed and draw solution were interchanged in PRO mode. All the membranes were tested in two different operational modes i.e., FO mode (active layer facing the feed solutions or AL-FS) and PRO mode (active layer facing the draw solution or AL-DS).

Initially, RO water was filled in both feed and the draw solution tank. Prior to experiments, RO water was circulated in FO mode for 15 min. Next, 2M NaCl solution replaced in draw solution tank and placed in digital weighing balance. The experiments were started after the flux stabilization and carried out for 90 min. The samples were analyzed at the time interval of 30 min. Again, water was replaced in draw solution and feed tank also was circulated for 15 min. Then, the membrane orientation towards PRO was conducted as per the above-mentioned procedure. The experiments were conducted in both FO and PRO modes. Power density (W/m$^3$) was calculated by following equation:

$$W = J_w\ \Delta P = A\ (\Delta\pi - \Delta P)\ \Delta P \qquad\qquad (2)$$

## Results and Discussion

Hydrophilicity is an important membrane parameter for having higher flux and more power density. Table 1 shows the contact angle values of TFC and TFN membranes. TFC membrane exhibits the higher contact angle value of 78.20°. It indicates that the water penetration rate was lower due to the hydrophobic character and higher surface tension. Lower contact angle value of 61.08° observed in 0.1 Zeolite 3Å incorporated TFN membrane. In TFN membranes, contact value gradually decreases with the increase of zeolite concentration up to 0.1 wt%. As zeolite incorporated TFN membrane encompasses rich hydrophilic silica and alumina groups. Thus, the addition of zeolite in the polyamide layer increases the water absorption rate. In PRO process, the higher hydrophilic membrane absorbs more water and results in the production of maximum power density. Although several recent findings indicated that the use of high amounts of nanoparticles (>1 wt %) during PA layer synthesis could lead to significant particle agglomeration in a selective layer which causes surface defects and reduced salt removal efficiency, Zeolite 3Å TFN membrane exhibits higher hydrophilic which is mainly due to the presence of higher content of $O_2$ in structure of zeolite.

**Table 2.** Contact angle, water permeability and salt rejection of TFC and zeolite incorporated TFN membranes

| Membrane Type | Contact Angle (°) | Water Permeability (L/m²h.bar) | Salt Flux (L/m²h) | Salt Rejection (%) |
|---|---|---|---|---|
| **TFC** | 78.20 | 0.51 | 6.23 | 96 |
| **0.05 Zt 3ÅTFN** | 69.3 | 0.59 | 8.19 | 95 |
| **0.1 Zt 3ÅTFN** | 61.08 | 0.64 | 11.37 | 94 |
| **0.3 Zt 3ÅTFN** | 64.88 | 0.61 | 9.55 | 94 |

## Conclusion

PRO is one of the most promising applications for power generation and membranes are the heart of this method. To improve this process, there are a variety of nanoparticles that are somehow promising to be incorporated with composite membranes. In addition, the recent results have shown that they are more promising to use in the PRO process in comparison with the control TFC membrane. Improving the performance of TFN membranes through PRO process is mostly rely on the nanomaterials using in the substrate or active layer (polyamide layer). It has been observed that using nanomaterials in TFN membranes has led into having higher water flux and consequently to have a higher power density and less fouling compared to the control TFC membrane. Working under different testing conditions will result in having various performance ranges which are between the 1.7 – 38.0 $W/m^2$. In addition, achieving better performance through the PRO process relies on the parameters such as different concentrations in the draw and feed solution and operation pressure. In this study, molecular sieves incorporated PA layer TFN membrane was fabricated successfully by an interfacial polymerization method. Molecular sieves have good compatibility with thin film PA layer. This result has good agreement with morphology analysis. The main required hydrophilic property of the membrane improved in molecular sieves incorporated TFN membranes. 0.1 Zeolite 3Å incorporated TFN membrane exhibits a lower contact angle and higher permeability of 61.08 and 0.64 $L/m^2h.bar$. Moreover, the higher water flux with better salt rejection was observed in 0.1 wt% molecular sieves incorporated TFN membranes. The optimal loading of molecular sieves is found to 0.1 wt%. Also, the higher water flux with minimal reverse solute flux was noticed in FO experiments for TFN membranes. The maximum flux of 5.6 $L/m^2h$ noticed in AL-DS mode for 0.1 Zeolite 3Å incorporated TFN membrane. This phenomenon is mainly due to molecular sieves, which improves the hydrophilicity of the membranes. Moreover, a higher power density of 1.73 $w/m^2$ observed in TFN membranes. These studies may be insight for other researchers and industrialist to the utilization of novel molecular sieves incorporated TFN membrane for desalination and membrane-based energy harvesting methods.

## References

1. Greenlee, Lauren F., Desmond F. Lawler, Benny D. Freeman, Benoit Marrot, and Philippe Moulin. (2009) "Reverse osmosis desalination: water sources, technology, and today's challenges." *Water research* 43, no. 9: 2317-2348.

2. Altaee, Ali, and Adel Sharif. (2015) "Pressure retarded osmosis: advancement in the process applications for power generation and desalination." *Desalination* 356: 31-46.

3. Zhang, Lizhi, Qianhong She, Rong Wang, Sunee Wongchitphimon, Yunfeng Chen, and Anthony G. Fane. (2016) "Unique roles of aminosilane in developing anti-fouling thin film composite (TFC) membranes for pressure retarded osmosis (PRO)." *Desalination* 389: 119-128.

4. Rad, Ali Shokuhi, Shadravan, Arvin, Soleymani, Amir Abbas and Motaghedi, Nazanin. (2015) "Lewis acid-base surface interaction of some boron compounds with N-doped graphene; first principles study." *Current Applied Physics* 15, no. 10: 1271-1277.

5. Achilli, Andrea, Tzahi Y. Cath, and Childress, Amy E. (2009) "Power generation with pressure retarded osmosis: An experimental and theoretical investigation." *Journal of membrane science* 343, no. 1-2: 42-52.

6. Lee, Esther Swin Hui, Jun Ying Xiong, Gang Han, Chun Feng Wan, Qing Yu Chong, and Tai-Shung Chung. (2017) "A pilot study on pressure retarded osmosis operation and effective cleaning strategies." *Desalination* 420: 273-282.

7. Han, Gang, Jieliang Zhou, Chunfeng Wan, Tianshi Yang, and Tai-Shung Chung. (2016) "Investigations of inorganic and organic fouling behaviors, antifouling and cleaning strategies for pressure retarded osmosis (PRO) membrane using seawater desalination brine and wastewater." *Water research* 103: 264-275.

8. Shadravan, Arvin, Pei Sean Goh, and Ahmad Fauzi Ismail. (2021) "Nanomaterials for Pressure Retarded Osmosis." In the book "*Advances in Water Desalination Technologies*", pp. 583-618.

9. Ma, Ning, Jing Wei, Rihong Liao, and Chuyang Y. Tang. (2012) "Zeolite-polyamide thin film nanocomposite membranes: Towards enhanced performance for forward osmosis." *Journal of Membrane Science* 405: 149-157.

10. Straub, Anthony P., Akshay Deshmukh, and Menachem Elimelech. (2016) "Pressure-retarded osmosis for power generation from salinity gradients: is it viable?" *Energy & Environmental Science* 9, no. 1: 31-48.

11. Li, Xue, Tao Cai, Gary Lee Amy, and Tai-Shung Chung. (2017) "Cleaning strategies and membrane flux recovery on anti-fouling membranes for pressure retarded osmosis." *Journal of Membrane Science* 522: 116-123.

12. Zhao, Die Ling, Subhabrata Das, and Tai-Shung Chung. (2017) "Carbon quantum dots grafted antifouling membranes for osmotic power generation via pressure-retarded osmosis process." *Environmental Science & Technology* 51, no. 23: 14016-14023.

13. Kim, H., J-S. Choi, and S. Lee. (2012) "Pressure retarded osmosis for energy production: membrane materials and operating conditions." *Water Science and Technology* 65, no. 10: 1789-1794.

14. Yip, Ngai Yin, Alberto Tiraferri, William A. Phillip, Jessica D. Schiffman, Laura A. Hoover, Yu Chang Kim, and Menachem Elimelech. (2011) "Thin-film composite pressure retarded osmosis membranes for sustainable power generation from salinity gradients." *Environmental Science & Technology* 45, no. 10: 4360-4369.

15. She, Qianhong, Xue Jin, and Chuyang Y. Tang. (2012) "Osmotic power production from salinity gradient resource by pressure retarded osmosis: Effects of operating conditions and reverse solute diffusion." *Journal of Membrane Science* 401: 262-273.

16. Achilli, Andrea, and Amy E. Childress. (2010) "Pressure retarded osmosis: from the vision of Sidney Loeb to the first prototype installation." Desalination 261, no. 3: 205-211.

17. Shadravan, Arvin, Goh, Peisean and Fauzi Ismail, Ahmad, (2019) "Thin film nanocomposite membranes incorporated with zeolite for power generation through pressure retarded osmosis", In the *14$^{th}$ International Conference on Membrane Science and Technology*, Nanyang Technological University (NTU), Singapore.

18. Shadravan, Arvin, Goh, Peisean and Fauzi Ismail, Ahmad, (2018) "Impact of Molecular Sieve Sizes on Tailoring Polyamide Layer Reverse Osmosis Membrane for Desalination Application", In the *National Congress on Membrane Technology*, Malaysia.

19. Shadravan, Arvin (2019) "Thin film nanocomposite membranes incorporated with zeolite for power generation through pressure retarded osmosis", *Master's Dissertation*, Universiti Teknologi Malaysia (UTM).

20. Yip, Ngai Yin, Alberto Tiraferri, William A. Phillip, Jessica D. Schiffman, Laura A. Hoover, Yu Chang Kim, and Menachem Elimelech. (2011) "Thin-film composite pressure retarded osmosis membranes for sustainable power generation from salinity gradients." *Environmental Science & Technology* 45, no. 10: 4360-4369.

21. Shokrolahzade Tehrani, Ali, Shadravan, Arvin, and Kashefi Asl, Morteza. (2017) "Investigation of Kinetics and Isotherms of Boron Adsorption of Water Samples by Natural Clinoptilolite and Clinoptilolite Modified with Sulfuric Acid." *Nashrieh Shimi va Mohandesi Shimi Iran* 35, no. 4: 21-32.

22. Inurria, Adam, Pinar Cay-Durgun, Douglas Rice, Haojie Zhang, Dong-Kyun Seo, Mary Laura Lind, and François Perreault. (2019) "Polyamide thin-film nanocomposite membranes with graphene oxide nanosheets: Balancing membrane performance and fouling propensity." *Desalination* 451: 139-147.